\documentclass[twocolumn]{aastex62}
\graphicspath{{./}{/}}

\submitjournal{ApJL}
\shorttitle{Gas Flows Within Cavities of Circumbinary Discs in Eccentric Binary Protostellar Systems}
\shortauthors{M\"osta et al.}

\begin{document}

\title[Gas Flows Within Cavities of Circumbinary Discs in Eccentric Binary Protostellar Systems]{Gas Flows Within Cavities of Circumbinary Discs in Eccentric Binary Protostellar Systems}

\correspondingauthor{Philipp M\"osta}
\email{pmoesta@berkeley.edu}

\author[0000-0002-0786-7307]{Philipp M\"osta} \affil{Department of Astronomy,
University of California at Berkeley, 501 Campbell Hall, Berkeley, CA 94720,
USA}

\author{Ronald E. Taam} \affiliation{Institute of Astronomy and Astrophysics -
TIARA, Academia Sinica, Taipei 10617, Taiwan} \affiliation{Center for
Interdisciplinary Exploration and Research in Astrophysics (CIERA), Department
of Physics and Astronomy,\\Northwestern University, 2145 Sheridan Road,
Evanston, IL 60208, USA} \author{Paul C. Duffell}
\affiliation{Harvard-Smithsonian Center for Astrophysics, 60 Garden Street,
Cambridge MA 02138, USA}



\begin{abstract}
The structure and evolution of gas flows within the cavity of a circumbinary
disk (CBD) surrounding the stellar components in eccentric binaries are
examined via two-dimensional hydrodynamical simulations. The degree to
which gas fills the cavity between the circumstellar disks (CSDs) and the
CBD is found to be greater for highly eccentric systems,
in comparison to low-eccentricity systems, reflecting the spatial
extent over which mass enters into the cavity throughout the orbit.
The pattern of the gas flow in the cavity differs for eccentric
binaries from that of binaries in a circular orbit.  In particular, the
former reveals tightly wound gas streams and figure-eight-like
structures for systems characterized by eccentricies, $e \geq 0.4$,
whereas the latter only reveal relatively loosely bent streams from the
CBD to the CSDs. Hence, the description
of the stream structures can be a probe of sufficient non-circularity
of the binary orbital motion.  Given that the inner edge of the
CBD is not very well defined for highly eccentric systems
due to the complex gas structures, it is suggested that the area of the
cavity for high-sensitivity imaging observations may prove to be a more
useful diagnostic for probing the effectiveness of CBD
clearing in the future. 
\end{abstract}


\section{Introduction}

Within the last decade radio  observations of protostellar binary systems at
submillimeter wavelengths using interferometers at the Submillimeter Array (SMA) 
and Atacama Large Millimeter/subillimeter Array (ALMA) have enabled
detailed spectral studies of their structures at high angular resolution.  Due
to the close proximity of protostellar binary systems in low mass star forming
regions ($< 200$ pc) and their large orbital separations ($\gtrsim 10$ AU) the
observations of L1551 NE (Takakuwa et al. 2014), UY Aur (Tang et al. 2014) and
GG Tau (Dutrey et al. 2016) have revealed significant morphological and
kinematic structure in the circumstellar disk (CSD) and circumbinary disk (CBD)
in these systems.  In particular, observations provide estimates of the shape
and size of the cavity within the CBD, the existence of gas streams from the
CBD to the CSDs, the stream connecting the CSDs, and shocks as inferred from
molecular tracers (e.g., SO).  All these diagnostic probes provide valuable
insight into the nature of these systems and the interactions between their
components. 

In parallel, the pioneering theoretical works of Artymowicz \& Lubow (1994,
1996) have provided an interpretative framework for some of these observational
data. In particular, the importance of the eccentricity of the orbit on the
inner size of the CBD was pointed out by Artymowicz \& Lubow (1994), Lubow \&
Artymowicz (1997) and summarized in a review by Dutrey et al. (2016).  Recent
numerical studies by Miranda, Munoz, \& Lai (2017), Thun, Kley, \& Picogna
(2017), and Munoz, Miranda, \& Lai (2018) and references therein have provided
theoretical understanding for the description of the structure within the CBD,
and the morphology and variability of the flows between the CBD and CSD.  Of
additional interest is the secular evolution of the mass accretion rates onto
the individual stellar components of the binary and the angular momentum
transfer rate between the CBD and the binary system.

In this Letter, we focus on the structure of the gas flowing from the CBD to the
CSDs within the cavity created by the non-central gravitational forces acting
on the gas within a binary system.  The general case of an eccentric binary is
considered as it is more appropriate for the systems under consideration where
the orbital separations of the observed protobinary systems with CBD are large.
Processes which can lead to the circularization of the orbital motion such as
tidal dissipation are ineffective.  

Previous studies by Gunther \& Kley (2002), Hanawa, Ochi, \& Ando (2010), and
de Val-Borro, Gahm, Stempels, \& Poplinksi (2011) have explored the temporal
and spatial properties of gas flows for specific binary star systems and,
hence, for a limited range of mass ratios and orbital eccentricities.  However,
we focus on the degree to which the cavity is filled and the resulting flow
patterns for a range of eccentricities and mass ratios.  We model the system
without excising the inner binary and CSDs and adopt a viscosity parameter
$\alpha = 0.001$, a value considered appropriate for disks surrounding young
stellar objects in contrast to higher values adopted in earlier studies.  Our
investigation is motivated, in part, by observational evidence for the presence
of shocked gas within the cavity of CBDs in binary protostars (e.g., UY Aur,
see Tang et al. 2014; L 1551 NE, see Takakuwa et al.  2017). As a major result,
we find that the flow patterns in the cavity can be characterized by either
tightly wound streams or figure-eight-like features, which are unique to
moderate and highly eccentric binary systems. 

In the next section, we briefly describe the numerical method, the assumptions
underlying our model, and the initial setup of our calculations.  The results
from a suite of two-dimensional hydrodynamical simulations are presented and
described with a particular focus on the patterns of the gas flow within the
cavity in \S 3.  The dependence of the character of the gas flow between the
CBD and CSDs on the eccentricity and mass ratio of the system are also
described.  Finally, we discuss the implications of our work, their relevance
to the interpretation of observations, and conclude in the last section. 

\begin{table} 
	\centering 
	\caption{Overview of parameters for simulations presented in this Letter.} 
	\begin{tabular}{cccc} \hline \hline Simulation & Mass Ratio q & Eccentricity e \\
	\hline
	q1.0e0.0 & 1.0 & 0.0 \\
	\hline
	q1.0e0.2 & 1.0 & 0.2 \\
	\hline
	q1.0e0.4 & 1.0 & 0.4 \\
	\hline
	q1.0e0.6 & 1.0 & 0.6 \\
	\hline
	q1.0e0.8 & 1.0 & 0.8 \\
	\hline
	q0.8e0.0 & 0.8 & 0.0 \\
	\hline
	q0.6e0.0 & 0.6 & 0.0 \\
	\hline
	q0.4e0.0 & 0.4 & 0.0 \\
	\hline
	q0.8e0.6 & 0.8 & 0.6 \\
	\hline
	q0.6e0.6 & 0.6 & 0.6 \\
	\hline
	q0.4e0.6 & 0.4 & 0.6 \\
	\hline
	\hline
        \end{tabular}
        \label{tab:simoverview}
        \vspace{0.2cm}
\end{table}

\section{Numerical Method and Assumptions}

\begin{figure*}
\centering
\includegraphics[width=0.32\textwidth]{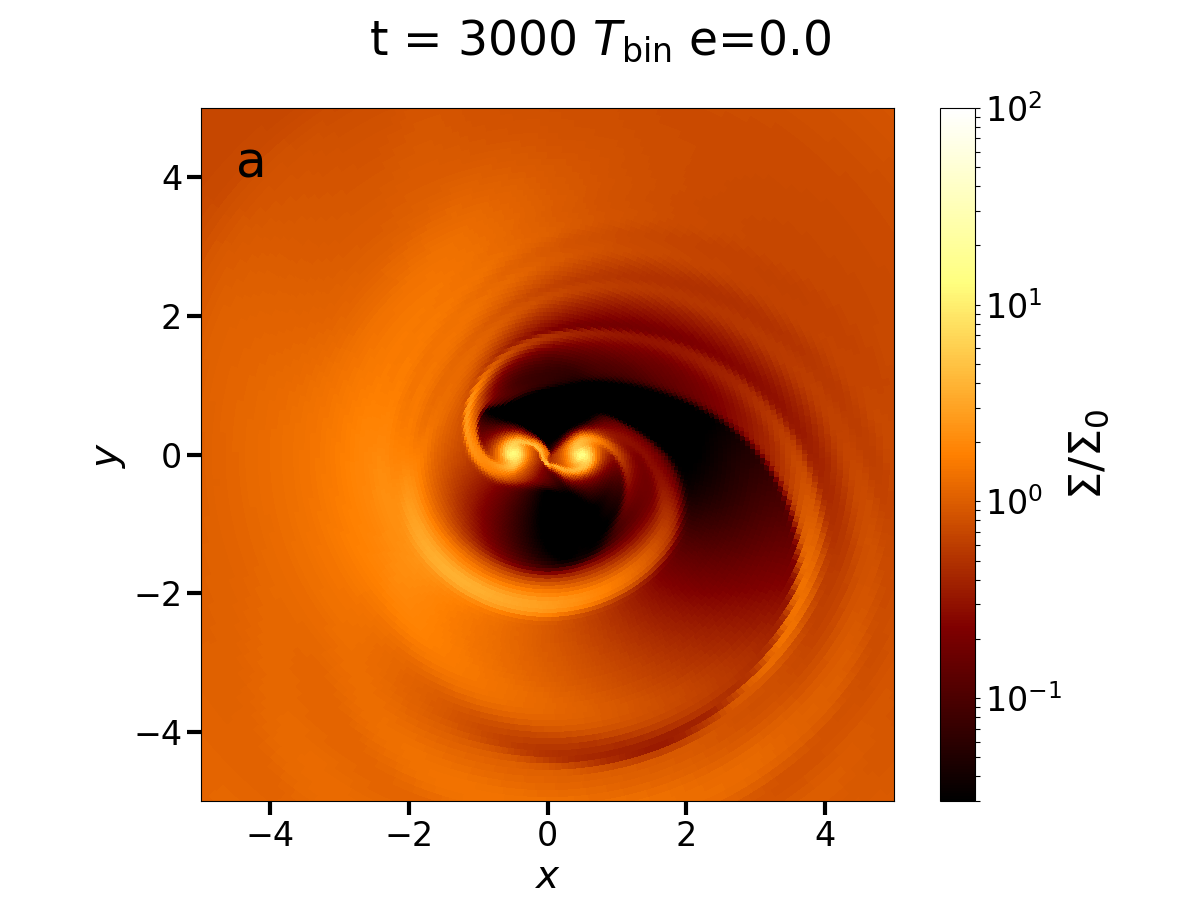}
\includegraphics[width=0.32\textwidth]{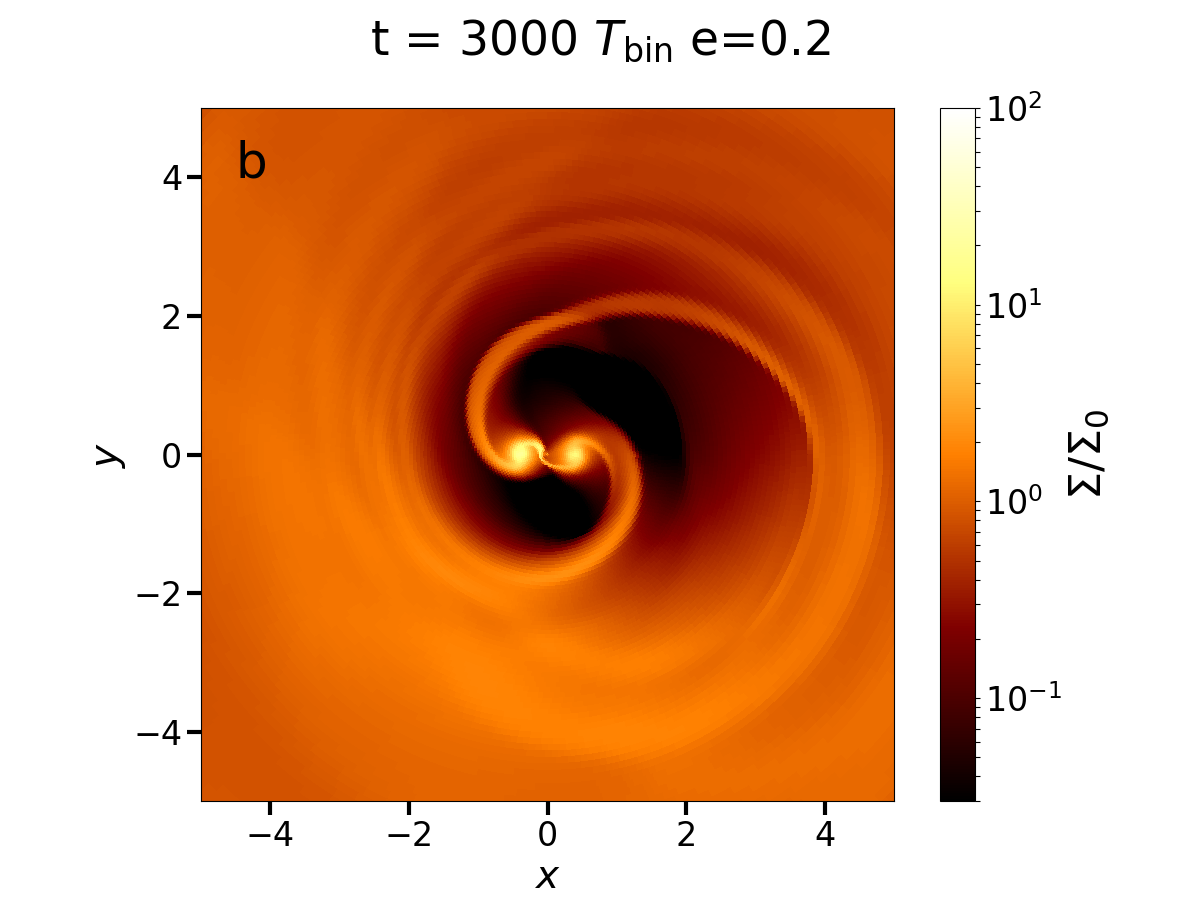}
\includegraphics[width=0.32\textwidth]{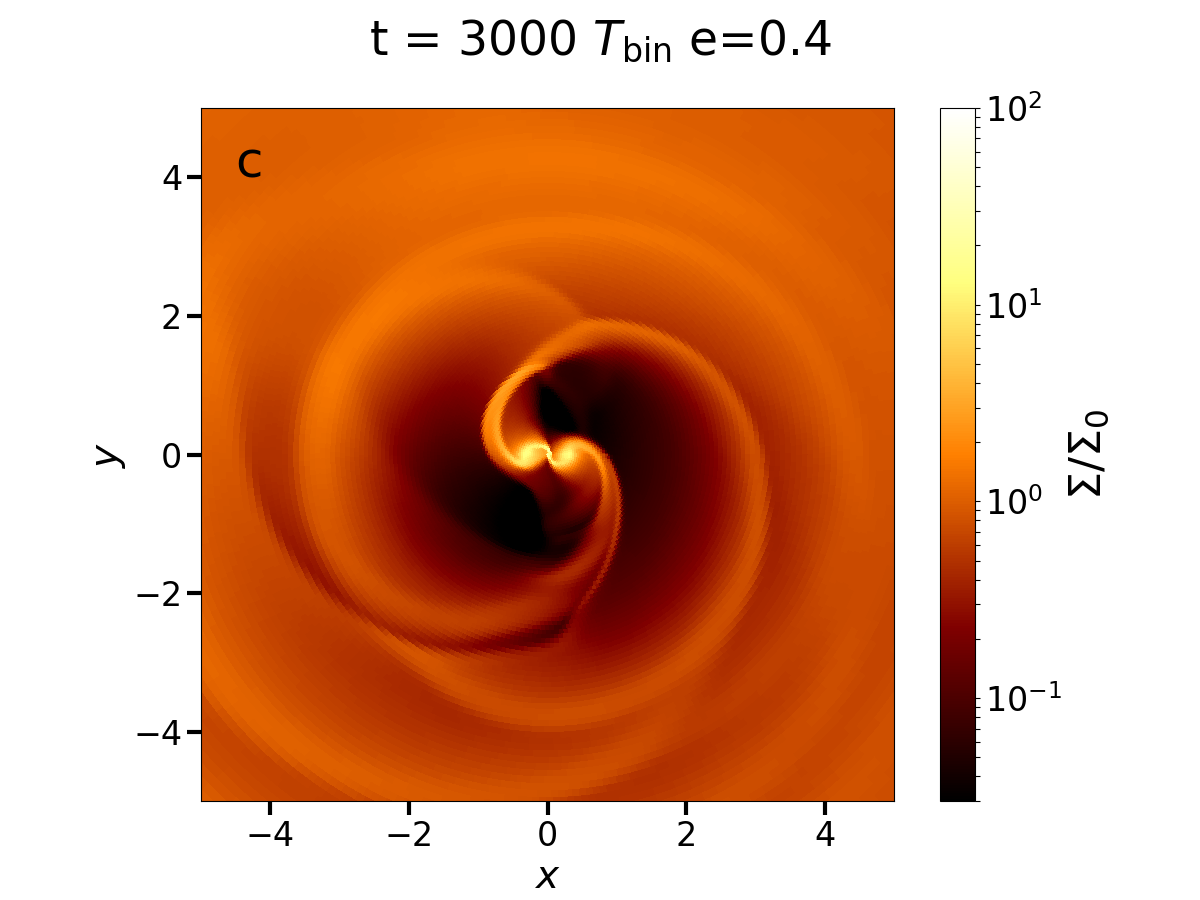}
\includegraphics[width=0.32\textwidth]{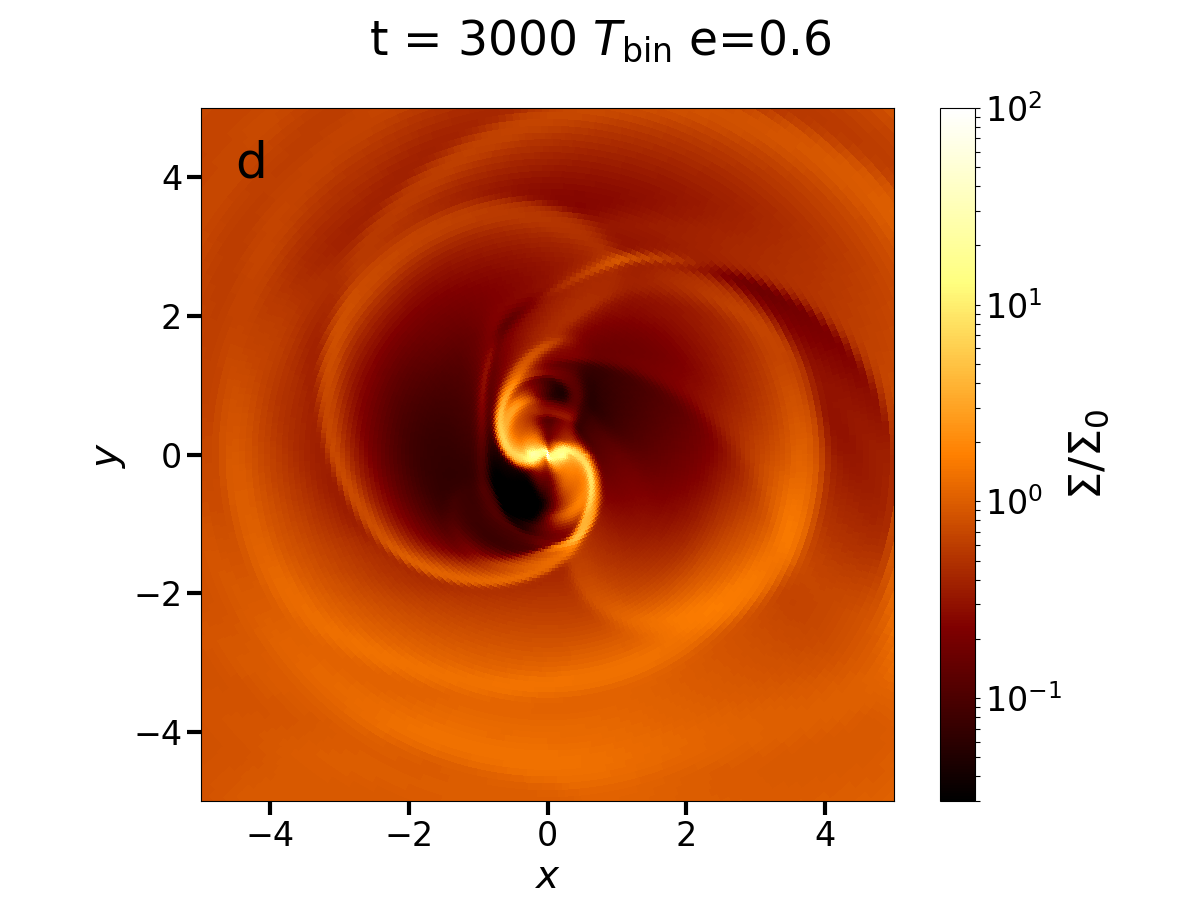}
\includegraphics[width=0.32\textwidth]{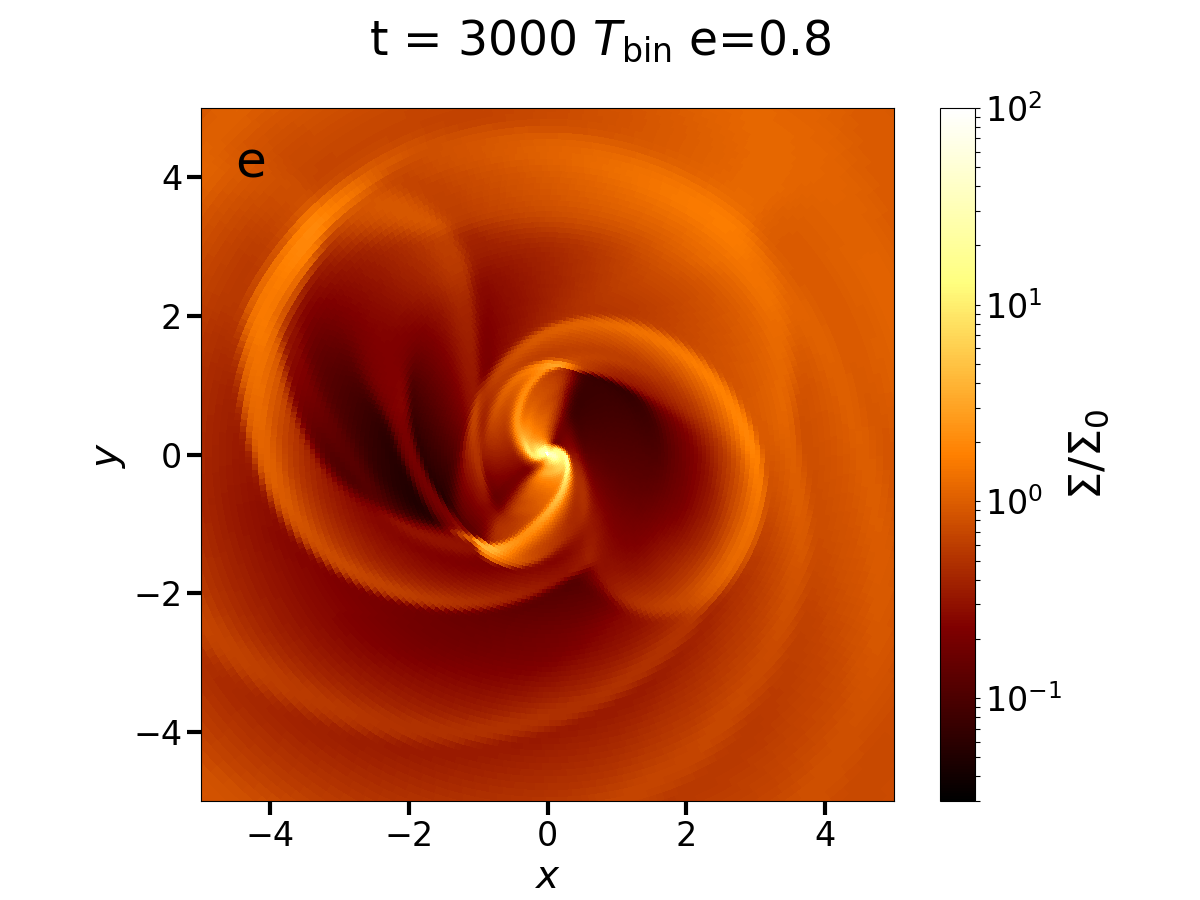}
	\caption{2D pseudocolor visualizations of surface density
	$\Sigma/\Sigma_0$ for the five equal-mass binaries q1.0e0.0 to
	q1.0e0.8. All panels show snapshots at $t = 3000\, T_{\mathrm{bin}}$. The
	color map is logarithmic and ranges from 0.03 to 100. The panels are
	zoomed in to show the region $x=y=\pm 5 a_{\mathrm{bin}}$ to highlight
	the behavior of the CSDs around the protostars and the
	CBD cavity. Panel (a) shows binary q1.0e0.0, panel (b) binary
	q1.0e0.2, panel (c) binary q1.0e0.4, panel (d) binary q1.0e0.6, and panel (e)
	binary q1.0e0.8.} 
\label{fig:2d_rho_zoom} 
\vspace{0.5cm} 
\end{figure*}

To model the hydrodynamical flow within the cavity of a CBD, we use the DISCO
code (Duffell \& MacFadyen 2012; Duffell 2016).  DISCO is a moving-mesh
hydrodynamics code specifically tailored to evolving astrophysical disks.
Computational zones are wedge-like annular segments that orbit with a
prescribed velocity.  Neighboring annuli shear with respect to each other such
that the mesh topology changes during the course of the calculation.  DISCO
evolves the hydrodynamic equations in cylindrical coordinates, explicitly
conserving angular momentum.

The mesh for our fiducial simulation setup consists of 128 annuli spaced from the
origin to $R = 10\, a_{\mathrm{bin}}$.  The highest resolution is attained near
the binary, with $\Delta R = 0.025 a_{\mathrm{bin}}$.  The azimuthal extent of
each zone is chosen such that $\Delta R \approx R \Delta \phi$, so that the
zones all have approximately aa  1:1 aspect ratio. The outer boundary at $R = 10\,
a_{\mathrm{bin}}$ fixes all fluid quantities to the initial conditions.  Zones
are constructed in such a way that there is no need for any special boundary
condition at $R=0$. We have also performed two simulations (q1.0e0.0 and
q1.0e0.6) at a factor 2 higher resolution of 256 annuli and in turn also
factor 2 higher resolution in azimuthal resolution. We find that the cavity shape
and area, as well the local flow patterns are only minimally changed when comparing
simulations with different resolution.

We carry out each numerical calculation for $5000$ orbits in order to achieve a
quasi-steady state in the inner region of the CBD and cavity region.  The inner
disk structure at late times is insensitive to the initial conditions.  The
binary is imposed as a pair of point masses orbiting one another with a fixed
mass ratio $q$, semi-major axis $a_{\mathrm{bin}}$ and eccentricity $e$.  The
binary motion is determined using the analytical solution to Kepler's
equations, so that the binary separation ranges from $a_{\mathrm{bin}}(1-e)$ to
$a_{\mathrm{bin}}(1+e)$.  The disk is setup by using an isothermal equation of
state, with $H/R = 0.2$ at $R=a_{\mathrm{bin}}$. We choose a uniform surface
density profile, $\Sigma(R) = \Sigma_0$, with $\Sigma_0 = 1.0$.  Viscosity is
chosen by imposing a fixed $\alpha = 0.001$, giving a kinematic viscosity of
$\nu = \alpha c_s^2 / \Omega$.

We use a sink prescription to model accretion onto the protostars from the
CSDs. We remove fluid smoothly from the vicinity of each protostar following
the prescription described in Farris et al. (2014). Specifically, we use a
source term $\left(\frac{d \Sigma}{dt}\right)_{\mathrm{sink,i}} = -
\frac{\Sigma}{t_{\mathrm{vis,i}}}$ which we add inside a radius
$r_{\mathrm{i}}i/a_{\mathrm{bin}} < 0.05$.  This removes fluid from the vicinity of each
protostar by assuming an $\alpha$-disk model for the CSD. We choose
$t_{\mathrm{vis,i}} = 0.1\, T_{\mathrm{bin}}$ to roughly match the expected
viscosity of the CSDs, but note that other choices ($t_{\mathrm{vis,i}} =
0.01\, T_{\mathrm{bin}}$ and $t_{\mathrm{vis,i}} = 0.001\, T_{\mathrm{bin}}$)
do not change the results and conclusions presented in this Letter qualitatively
(as noted by Tang et al 2018, this choice can affect the torque measured, but
that is not the focus of this study).

\section{Simulation Results} We have performed simulations at low viscosity
($\alpha=10^{-3}$) and for a sink rate of $0.1$ for different mass ratios and
eccentricities. The parameters and key quantities from the simulations are
presented in Table \ref{tab:simoverview}. We have explored mass ratios between
1.0 and 0.4 and eccentricities between $e=0.0$ and $e=0.8$. 
The equal-mass circular orbits simulation q1.0e0.0 is our reference simulation.

We find differences in flow structure between the CBD and the two individual
protostellar disks that surround each of the stars. In the following we first
describe the dependence of the flow structure on the eccentricity of the binary
stars' orbital motion and secondly describe the dependence of the accretion
flows onto the two protostellar disks when the two stars have different masses.
\\\\

2D pseudocolor visualizations of surface density for the simulation series with
varying eccentricities (q1.0e0.0 to q1.0e0.8) are shown in
Fig.~\ref{fig:2d_rho_zoom}. To highlight the behavior of the binary and disk
cavity we limit the extension of the color plots to $x=y=\pm 5
a_{\mathrm{bin}}$. The CBD cavity becomes eccentric in all cases, but the
cavity size is larger in radial extent for higher eccentricities of the binary.
At the same time, the material inside the cavity has higher density and 
its mass is greater for higher eccentricities. Torques
exerted by the binary impact the disk structure out to larger radii for 
more eccentric orbits. The CSDs surrounding the individual protostars are more
extended in the case of zero or low eccentricity. In contrast, the accretion
streams onto the CSDs from the CBD are more diffuse and less clearly defined
for higher eccentricities.

\begin{figure*}
\centering
\includegraphics[width=0.28\textwidth]{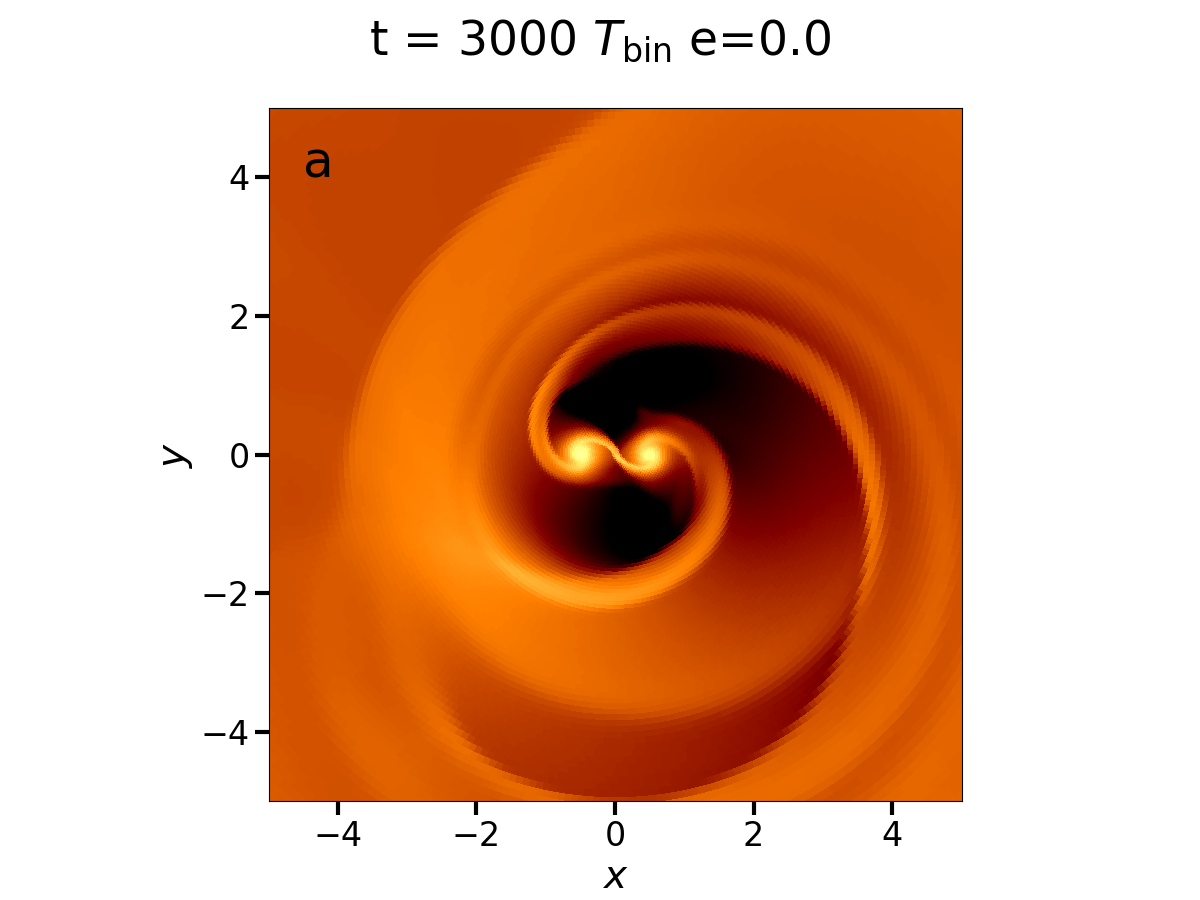}
\hspace{-1.2cm}
\includegraphics[width=0.28\textwidth]{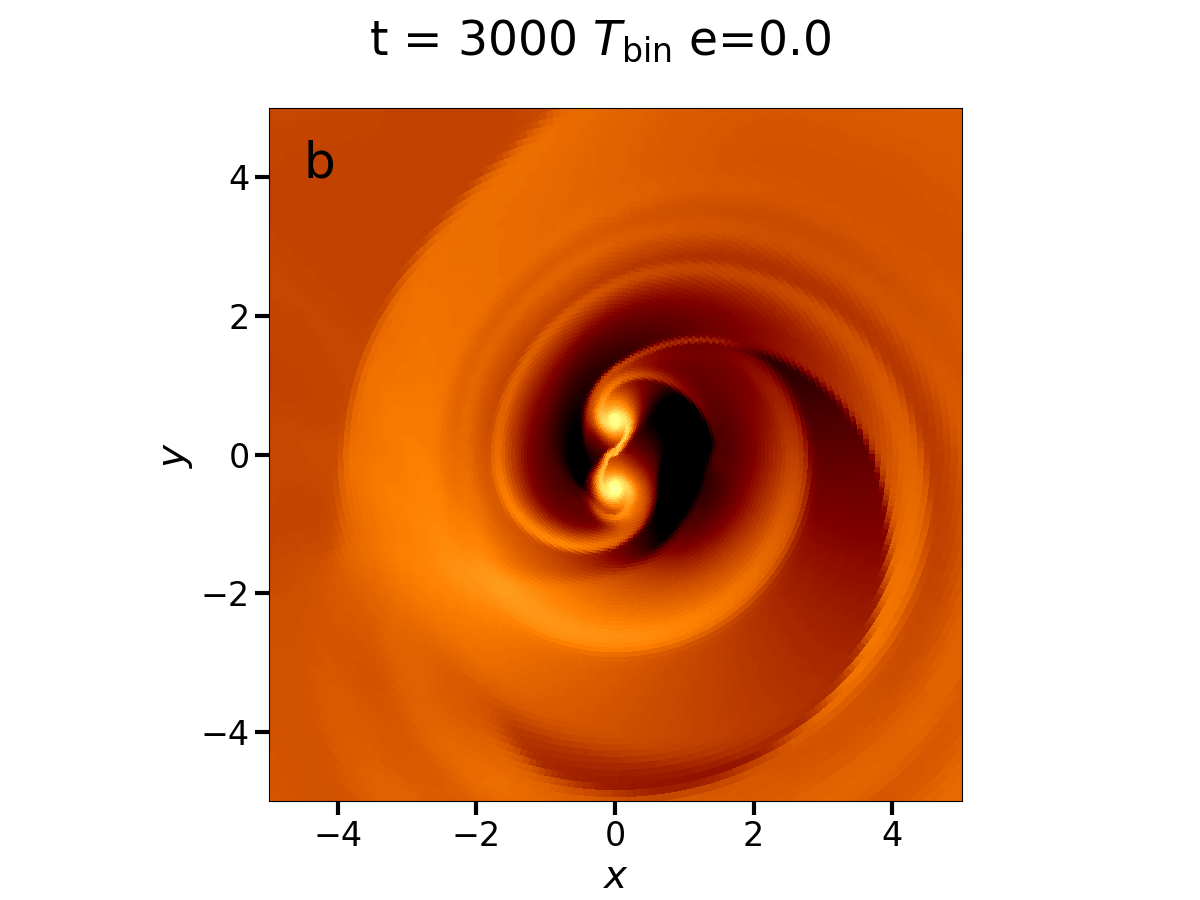}
\hspace{-1.2cm}
\includegraphics[width=0.28\textwidth]{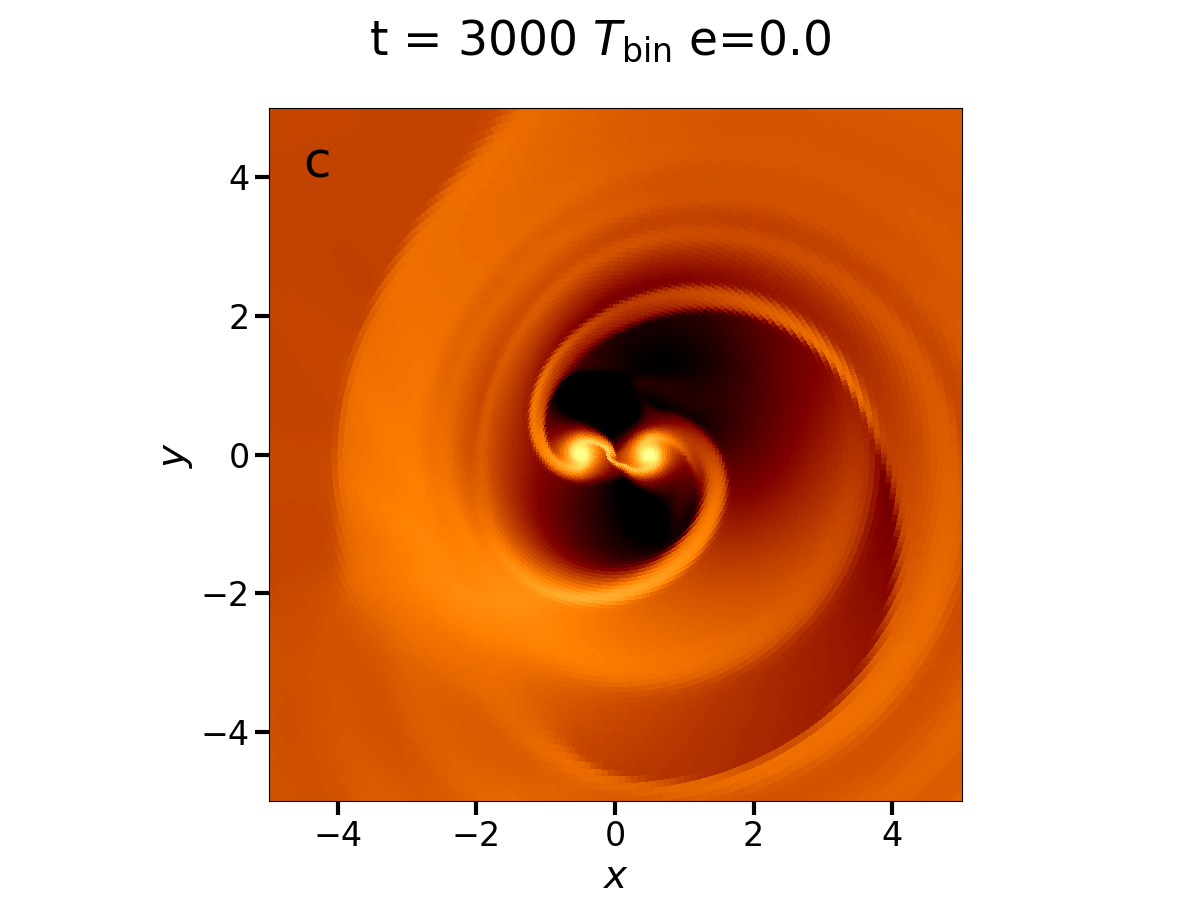}
\hspace{-0.95cm}
\includegraphics[width=0.28\textwidth]{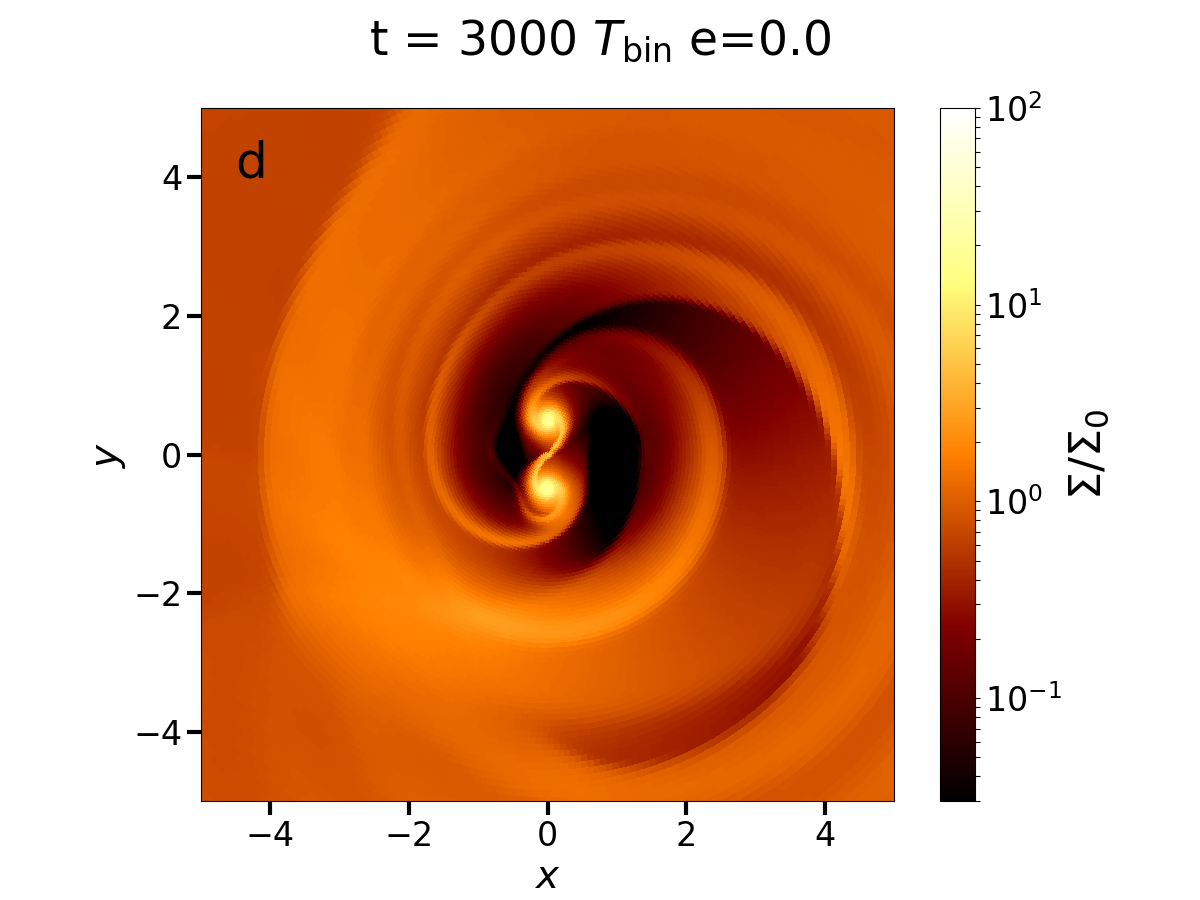}
\includegraphics[width=0.28\textwidth]{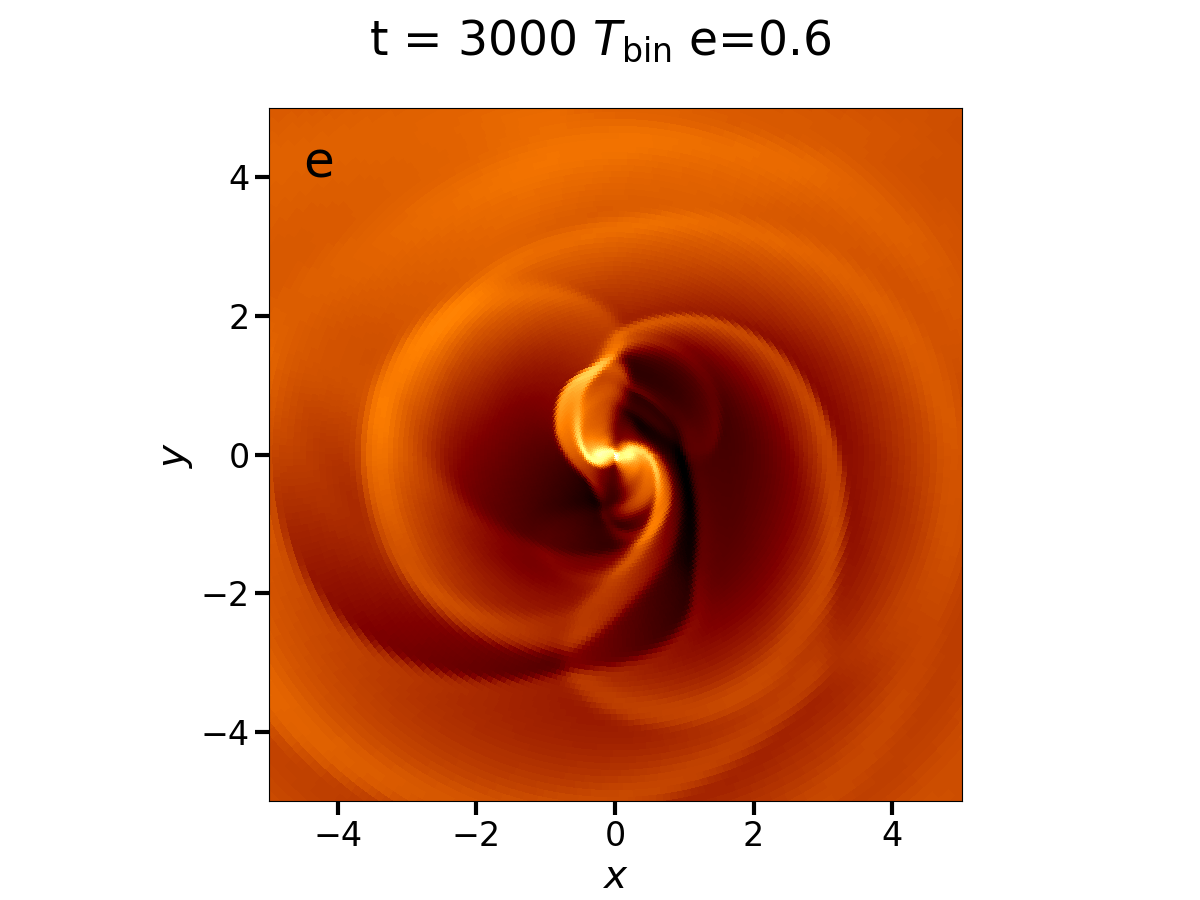}
\hspace{-1.2cm}
\includegraphics[width=0.28\textwidth]{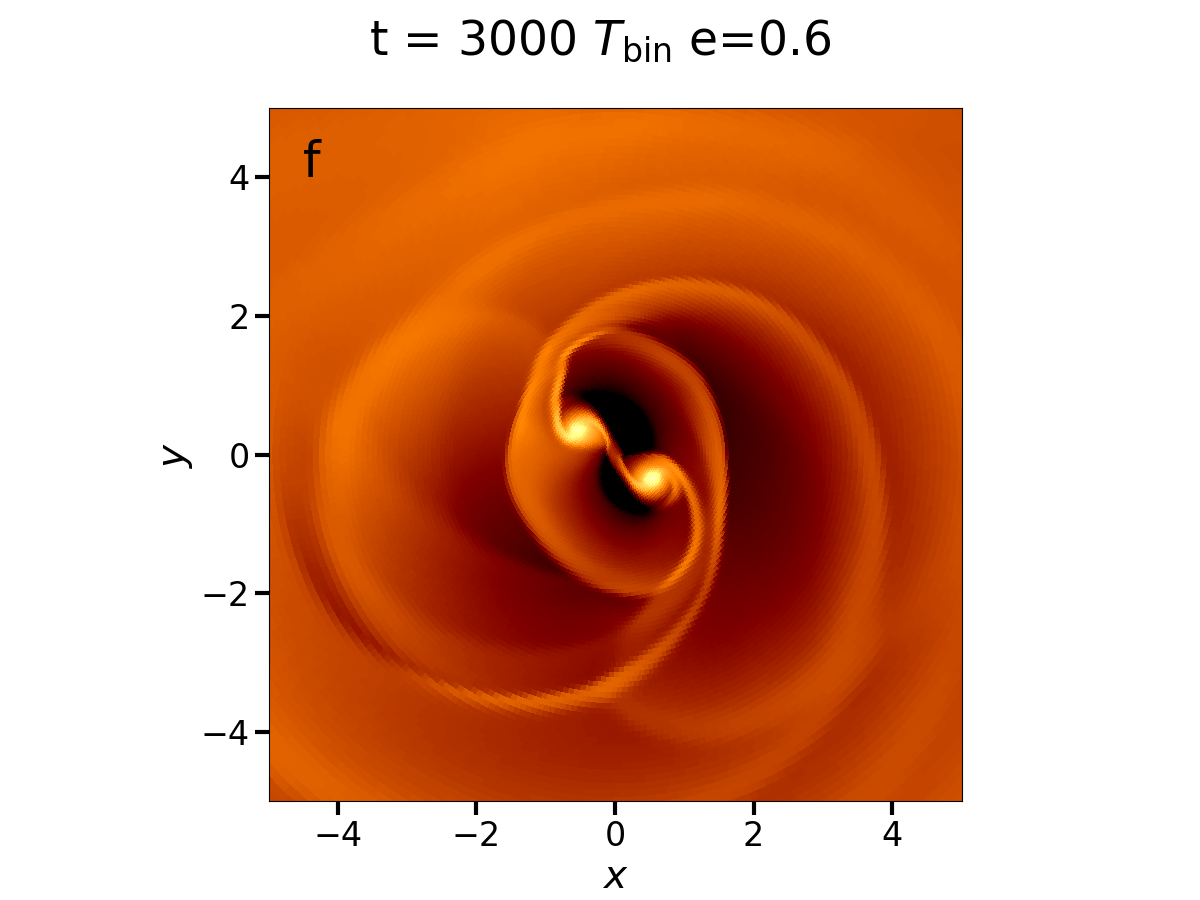}
\hspace{-1.2cm}
\includegraphics[width=0.28\textwidth]{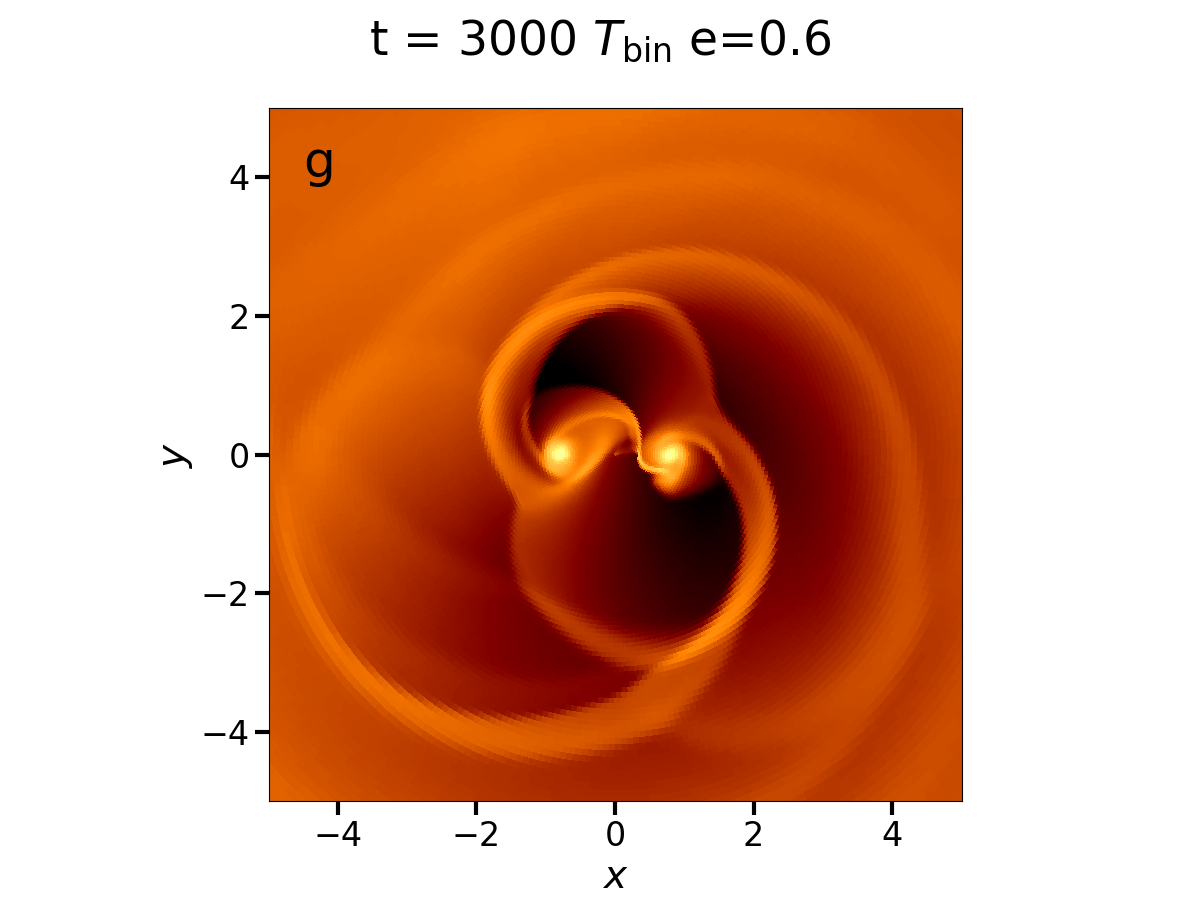}
\hspace{-0.95cm}
\includegraphics[width=0.28\textwidth]{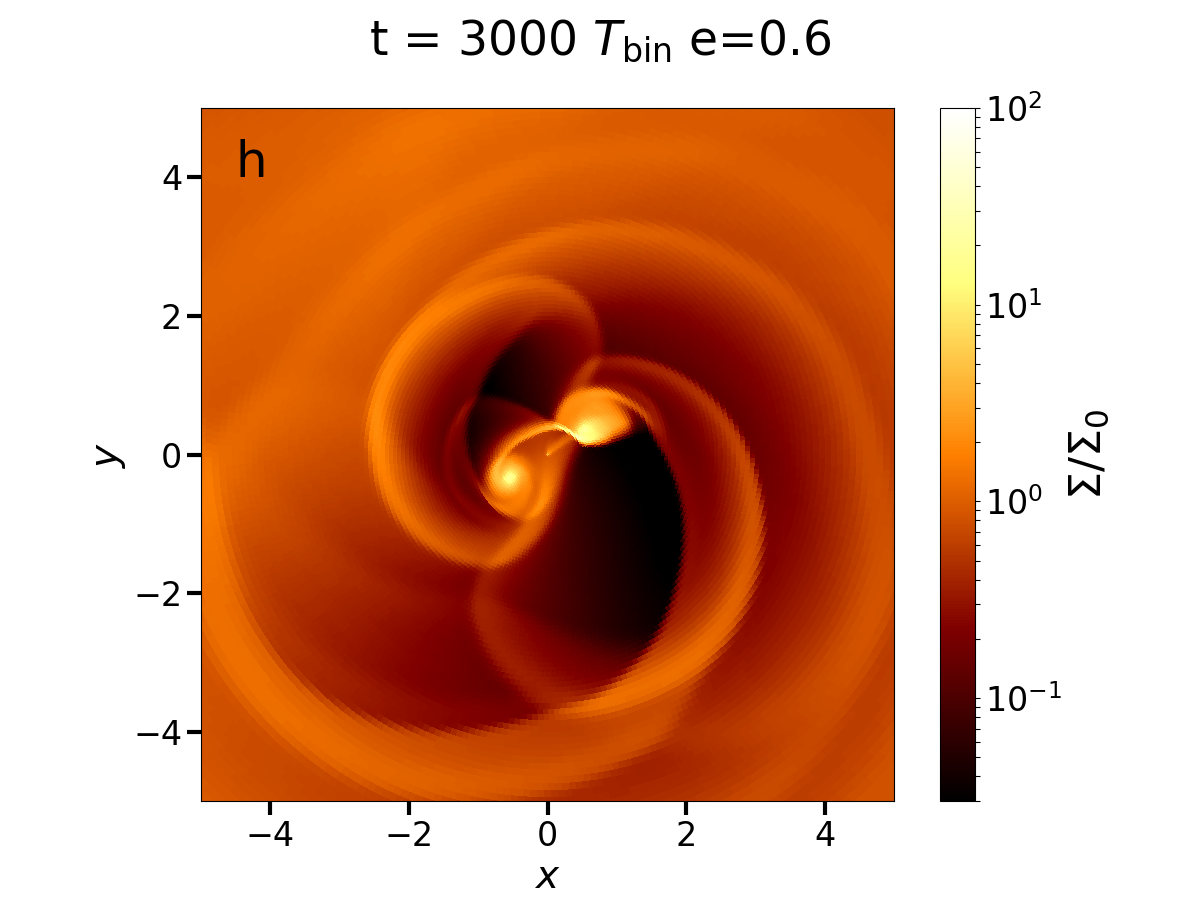}
\includegraphics[width=0.28\textwidth]{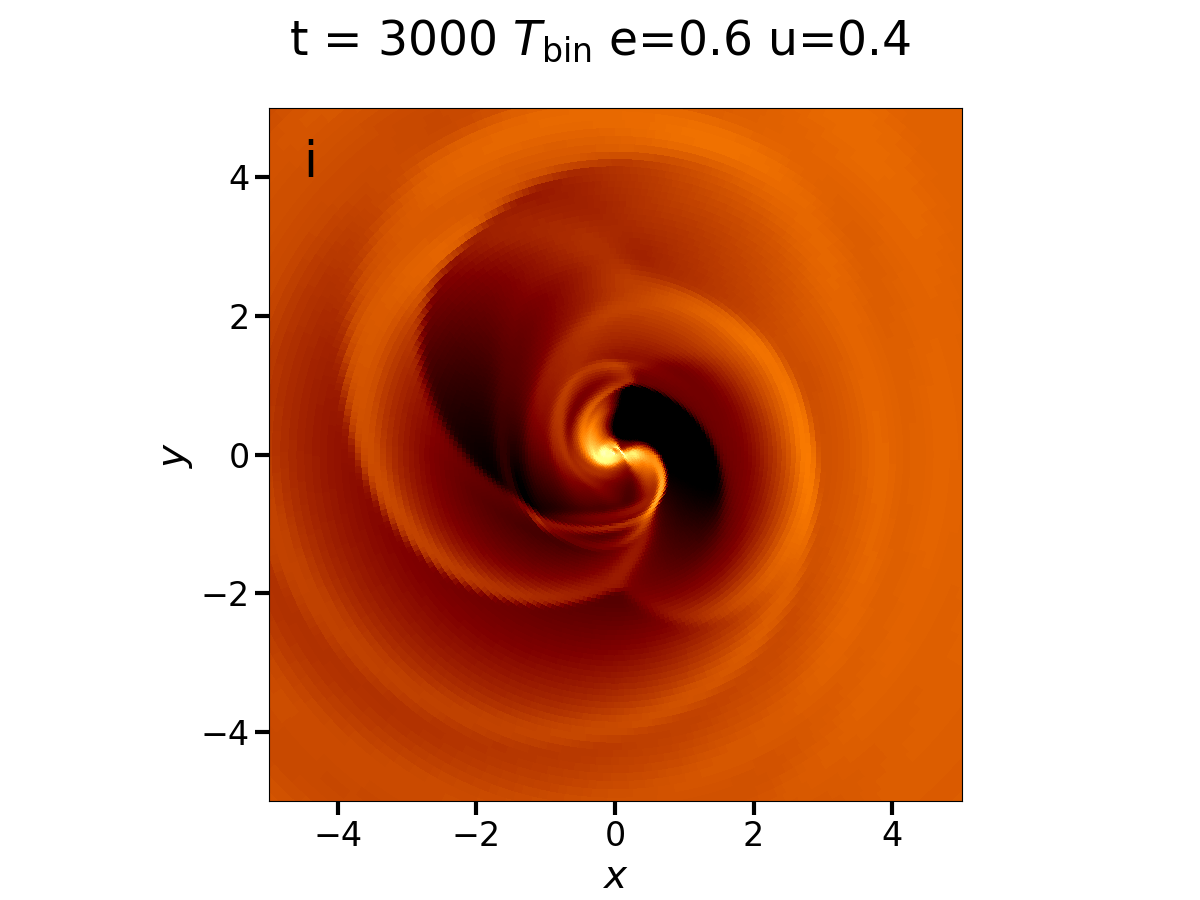}
\hspace{-1.2cm}
\includegraphics[width=0.28\textwidth]{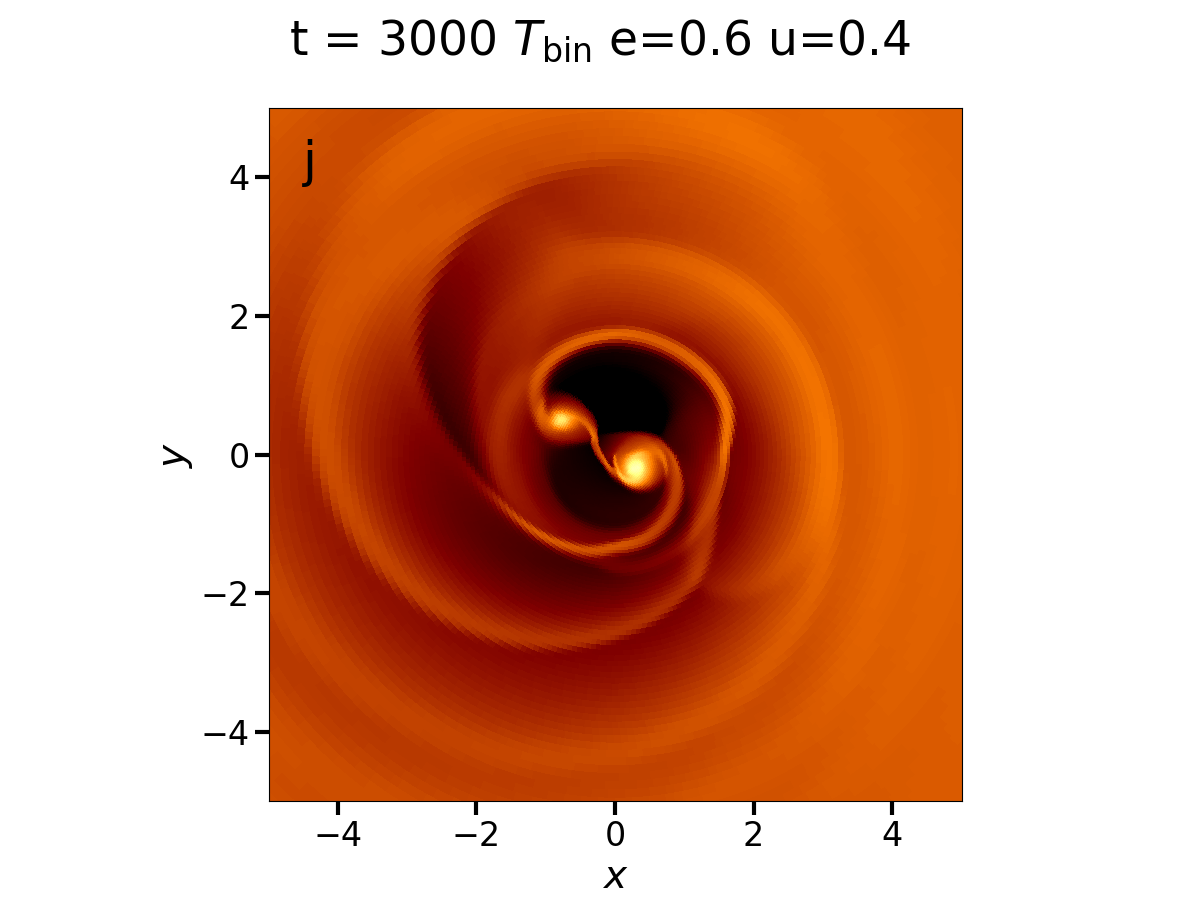}
\hspace{-1.2cm}
\includegraphics[width=0.28\textwidth]{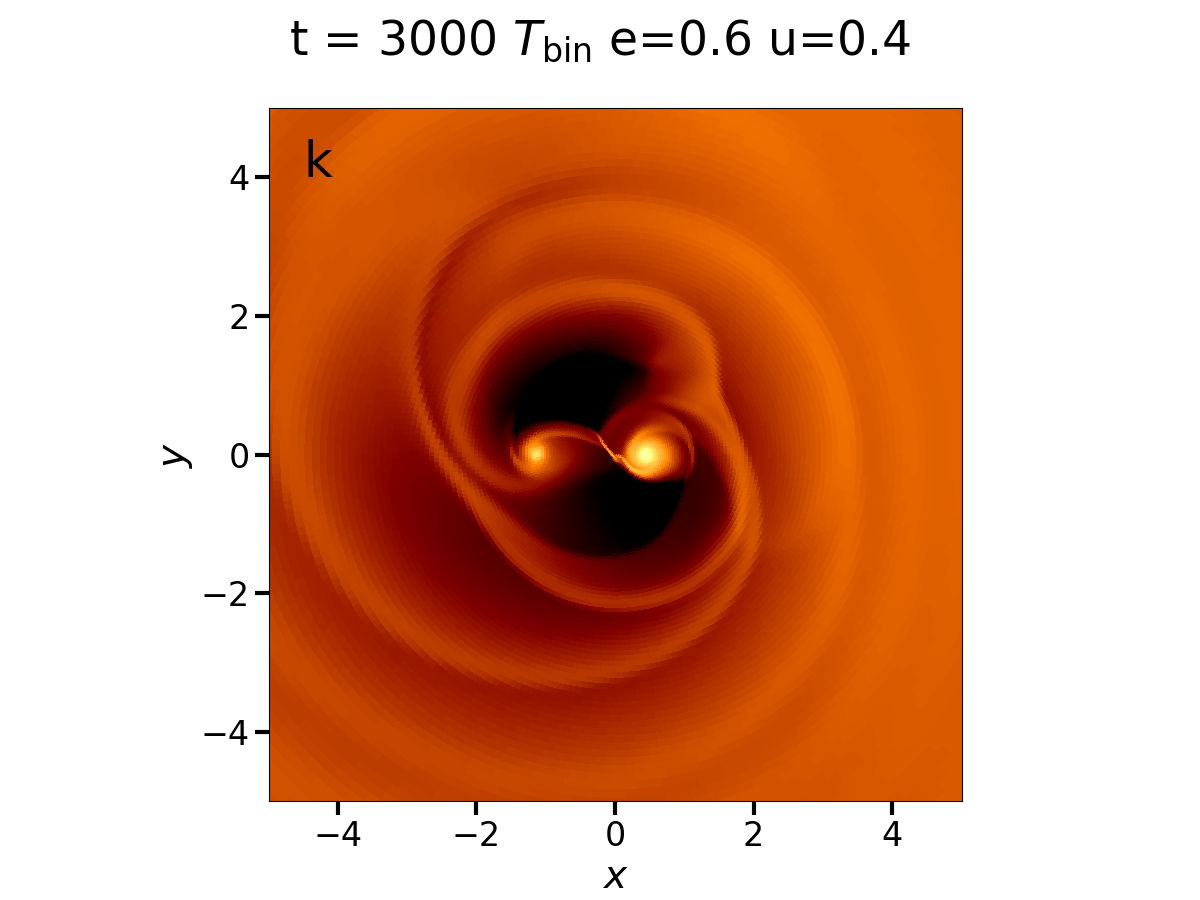}
\hspace{-0.95cm}
\includegraphics[width=0.28\textwidth]{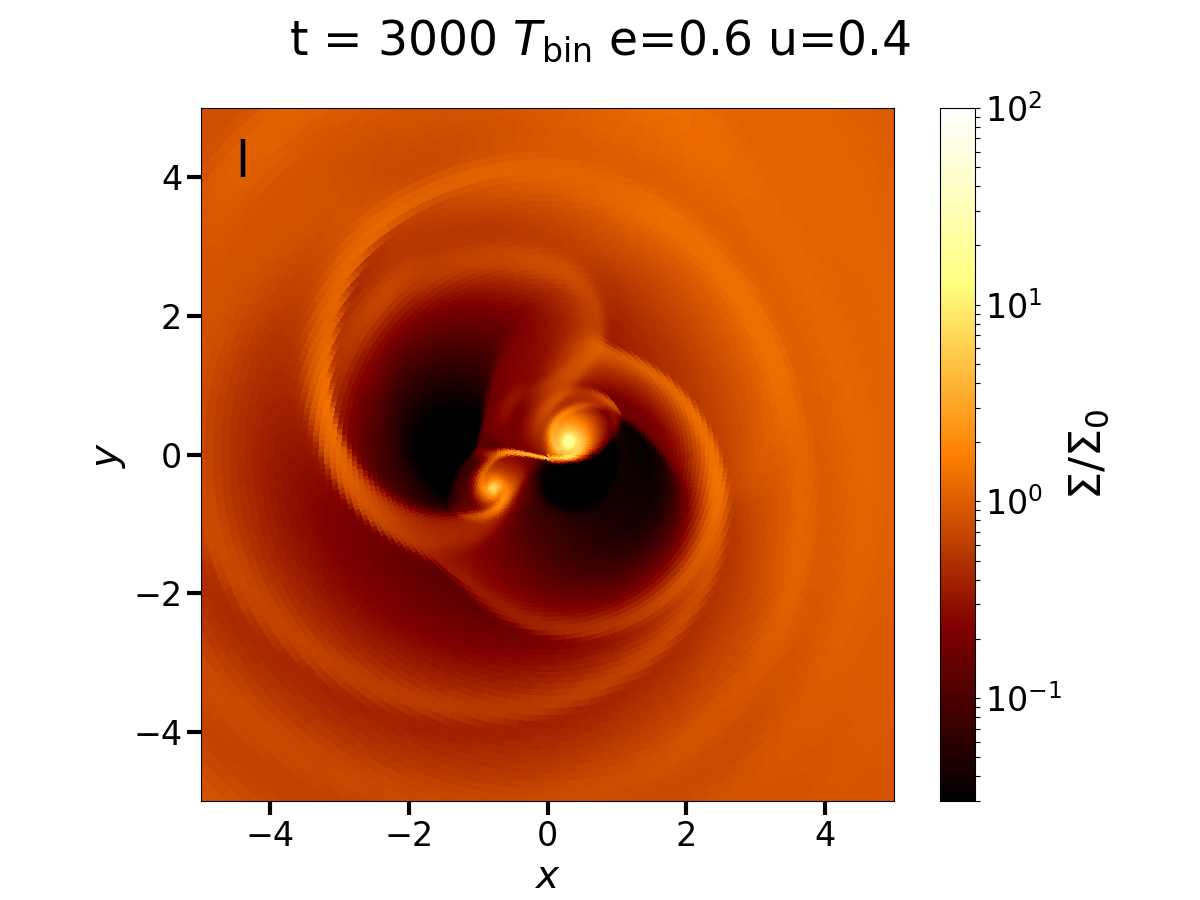}

\caption{2D pseudocolor visualizations of surface density $\Sigma/\Sigma_0$ for the 
	equal-mass binary q1.0e0.0 (top row, panels (a)-(d)), the equal-mass, eccentric
	binary q1.0e0.6 (middle row, panels (e)-(h)), and the unequal-mass, eccentric
	binary q0.4e0.6 (bottom row, panels (i)-(l)) during different phases 
	of a single orbit. Panels (a)-(d) show four snapshots during the orbit
	starting $t = 3000\, T_{\mathrm{bin}}$ for q1.0e0.0. Panels (e)/(i) show
	the eccentric binaries q1.0e0.6/q0.4e0.6, respectively, at periastron,
	panels (f)/(j) in between periastron and apastron, panels (g)/(k) at apastron,
	and panels (h)/(l) again in between apastron and periastron. The color map
	is logarithmic and ranges from 0.03 to 100. The panels are zoomed in to
	show the region $x=y=\pm 5 a_{\mathrm{bin}}$ to highlight the behavior
	of the CSDs around the protostars and the CBD
	cavity.} 
\label{fig:2d_rho_1orbit} 
\vspace{0.5cm} 
\end{figure*}

To illustrate the differences in the flow structure in the orbital evolution of
the binary and CBD, we show snapshots at different times during a single orbit
for simulations q1.0e0.0, q1.0e0.6, and q0.4e0.6.  Panels (a)-(d) of
Fig.~\ref{fig:2d_rho_1orbit} illustrate that the differences in flow structure
during one orbit vary only moderately in the case for a circular orbit. Both
the two CSDs around the protostars and the accretion streams onto them remain
well connected and defined during the orbit. A stream connecting the two CSDs
is always present during the orbit. For comparison we show the eccentric binary
q1.0e0.6 in panels (e)-(h) of Fig.~\ref{fig:2d_rho_1orbit}. Here, the orbital
dynamics and changes in binary separation during one orbit change the behavior
of the accretion flows, CSDs and stream between the CSDs significantly. At
periastron (Fig.~\ref{fig:2d_rho_1orbit} panel (e), the two CSDs nearly touch as
the orbital separation is the smallest. The accretion streams onto the CSDs are
relatively narrow and well defined. As the binary evolves toward apastron
(Fig.~\ref{fig:2d_rho_1orbit} panels (f),(g)), the accretion streams become more
diffuse and the cavity structure more complicated. A figure-eight-like structure
emerges that is not present in the circular orbit case. The flow between the
two CSDs becomes very narrow and less dense than in the circular orbit case. In
the evolution back toward periastron (Fig.~\ref{fig:2d_rho_1orbit} panel (h)),
the structure disappears again. Finally, panels (i)-(l) of
Fig.~\ref{fig:2d_rho_1orbit} show the eccentric unequal-mass binary q0.4e0.6.
The flow patterns are very similar to the equal-mass, eccentric binary q1.0e0.6
and the same figure-eight-like structure emerges in the evolution toward apastron.
This structure is also present in the binaries q1.0e0.4 and
q1.0e0.8 (not shown here).  Overall, in all simulations with eccentricity
$e\geq 0.4$, there is more diffuse low-density gas in the cavity during all
stages of the orbit compared to the circular orbit case and the figure-eight-like
structure in the accretion flows onto the CSDs appears as a robust feature of
the eccentricity in the orbits of the two stars. 

\begin{figure*}
\centering
\includegraphics[width=0.47125\textwidth]{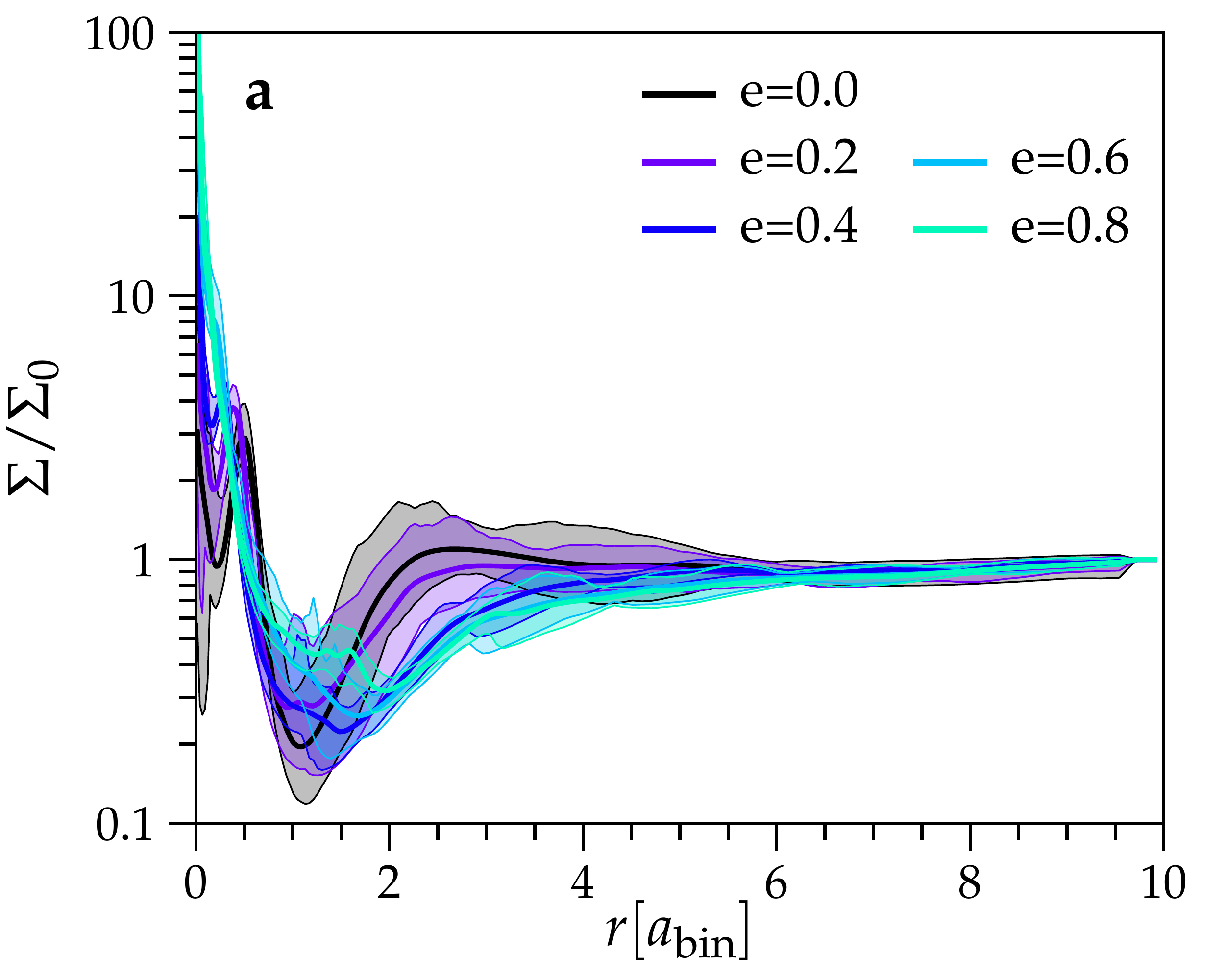}
\includegraphics[width=0.47125\textwidth]{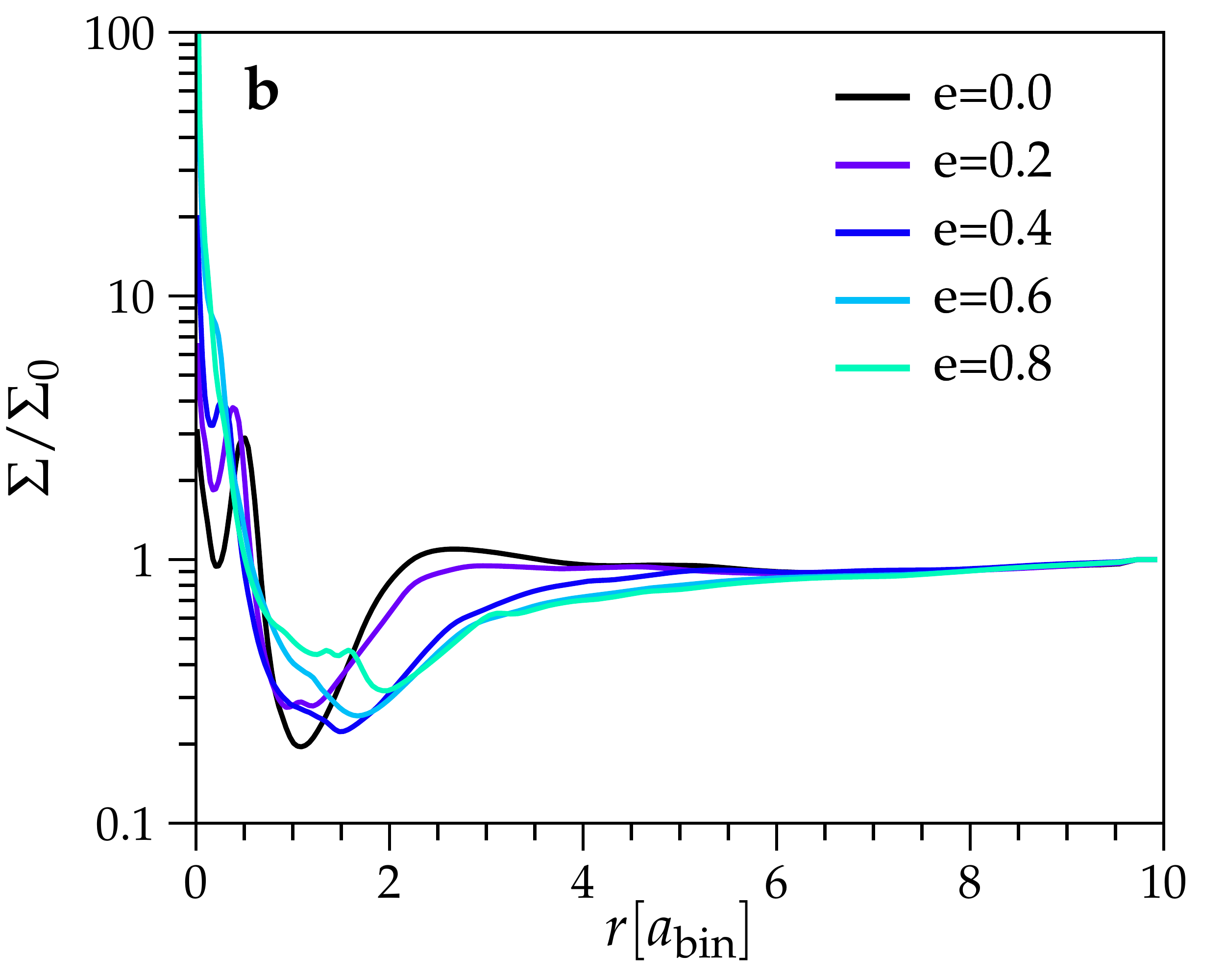}
\caption{Azimuthally averaged surface density as a function of radius for the equal-mass binaries
	q1.0e0.0 - q1.0e0.8. The black lines depict the binary
	q1.0e0.0, the purple line the q1.0e0.2 binary, the blue line the
	q1.0e0.4 binary, the light blue line the q1.0e0.6 binary, and the cyan
	line the q1.0e0.8 binary. Panel (a) shows the mean surface density
	profile over the last 3000 orbits ($2000-5000\, T_{\mathrm{bin}}$). The
	shaded regions indicate the maximum and minimum values of all
	individual profiles considered in the mean. Panel (b) shows the mean profiles
	only to indicate the differences in cavity size and filling.}
\label{fig:rho_vs_r} 
\vspace{0.5cm} 
\end{figure*}

\begin{figure}
\centering
\includegraphics[width=0.47125\textwidth]{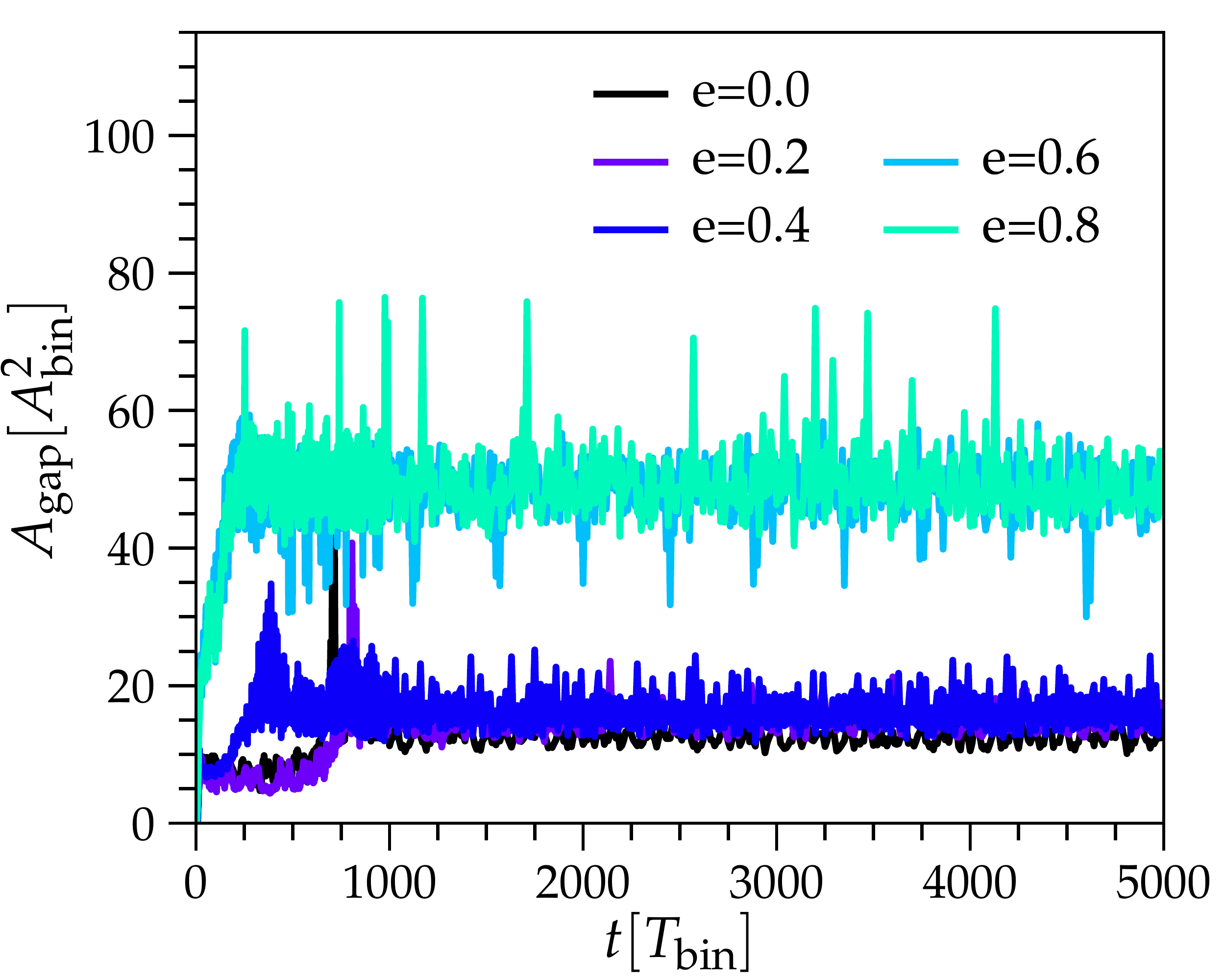} \caption{
	Area of the CBD cavity as a function of time for the
	equal-mass binaries q1.0e0.0 - q1.0e0.8. We define the area of the
	cavity as the area of all zones below a surface density of
	$\Sigma/\Sigma_{\mathrm{CSD,max}} = 0.025$, where 
	$\Sigma_{\mathrm{CSD,max}}$ is the maximum density in the CSD. 
	The black line depicts the area for the binary q1.0e0.0, the purple
	line the q1.0e0.2 binary, the blue line the q1.0e0.4 binary, the light
	blue line the q1.0e0.6 binary, and the cyan line the q1.0e0.8 binary.}
\label{fig:area_vs_t} 
\vspace{0.5cm} 
\end{figure}

Fig.~\ref{fig:rho_vs_r} shows angle-averaged (in azimuthal angle $\phi$)
profiles of surface density $\Sigma / \Sigma_0$ as a function of radius $r$ for
the binaries q1.0e0.0 - q1.0e0.8. Panel (a) (left) shows the time-averaged
surface density profiles for $ 2000\, T_{\mathrm{bin}} \leq t \leq 5000\,
T_{\mathrm{bin}}$ (solid lines). The shaded regions surrounding the
time-averaged profiles show the minimum and maximum values of profiles at
individual times and give an indication of the variability of the radial
structure over time. Panel (b) (right) shows only the time-averaged surface
density profiles as a function of radius.  It can be seen that clear trends
with eccentricity are present. The radius of the inner edge of the CBD becomes
larger for increasing eccentricity. At the same time the cavity remains filled
with higher-density material compared to the circular orbit case, making it
difficult to define the location of the inner edge of the CBD with increasing
eccentricity. In addition there is less variation around the mean profiles for
increasing eccentricity. We note that due to the chosen value for the viscosity
parameter of $\alpha = 10^{-3}$, only the cavity and inner regions of the CBD
evolve for a full viscous time at $t = 5000\, T_{\mathrm{bin}}$, but there is
still little to no evolution of the outer regions of the CBD. This is directly
visible in the very small extent of the shaded regions beyond $r \simeq 5\,
a_{\mathrm{bin}}$. This however does not impact on the results of this Letter as
our focus is exclusively on the behavior of the cavity surrounding the binary
and the accretion flows onto and between the CSDs surrounding the stars.  As a
final remark, we point out that the surface density of material in the cavity
does not decrease as much in our simulations when compared to simulations that
excise the stars and CSDs surrounding them (see, e.g., Miranda, Munoz, \& Lai
2017 and Thun, Kley, \& Picogna 2017) due to the absence of an imposed inner
boundary.

\begin{figure*}
\centering
\includegraphics[width=0.45\textwidth]{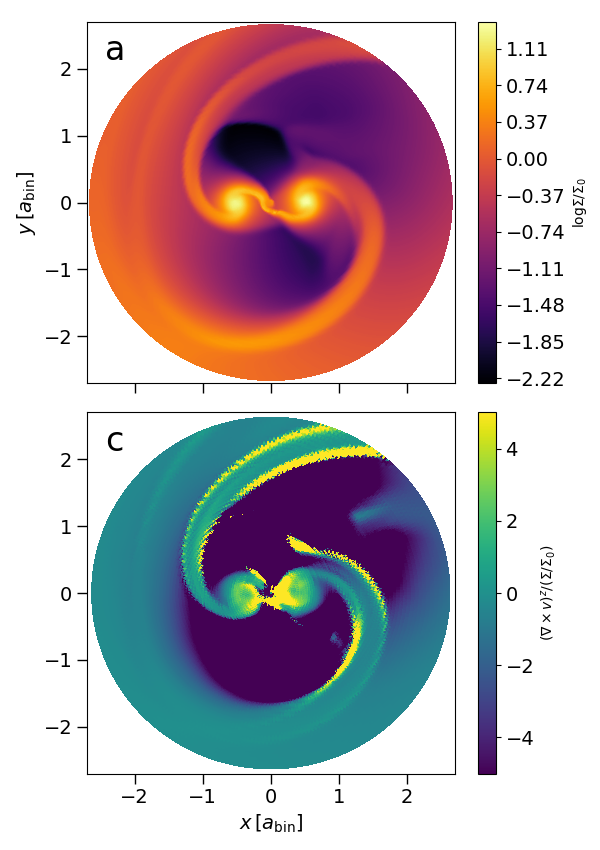}
\includegraphics[width=0.45\textwidth]{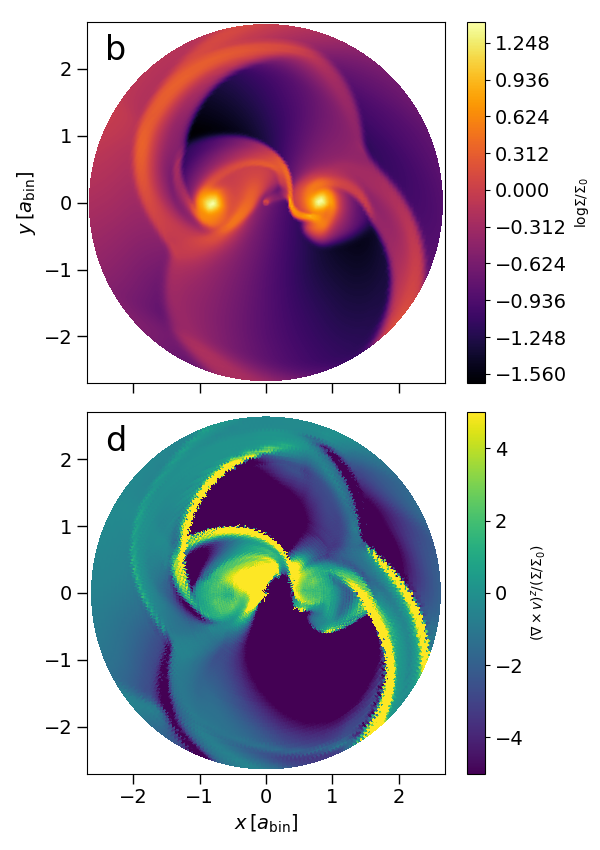}
	\caption{2D pseudocolor visualizations of surface density
	$\Sigma/\Sigma_0$ (top row) and the z-component of the potential vorticity
	$\left(\nabla \times \vec{v}\right)^z / (\Sigma/\Sigma_0)$ (bottom row)
	for the equal-mass binary q1.0e0.0 (left column, panels (a),(c)) and the
	eccentric binary q1.0e0.6 (right column, panels (b),(d)) at periastron at
	$T = 3000\, T_{\mathrm{bin}}$. The color map is logarithmic for
	$\Sigma/\Sigma_0$ and linear for $\left(\nabla \times \vec{v}\right)^z /
	(\Sigma/\Sigma_0)$. The panels are zoomed in to show the region
	$x=y=\pm 2.7 a_{\mathrm{bin}}$ to highlight the behavior of the CSDs
	around the protostars and the CBD cavity. The numerical noise in the
	center of panels (c) and (d) is due to interpolation from the
	\texttt{Disco} mesh to a regular spaced mesh to evaluate $\left(\nabla \times
	\vec{v}\right)^z$ via finite difference.} 
\label{fig:2d_divv} 
\vspace{0.5cm} 
\end{figure*}

To quantify the differences in the description of the cavity as a
function of eccentricity we define the area of the cavity as the area of all
zones below a surface density of $\Sigma/\Sigma_{\mathrm{CSD,max}} = 0.025$,
where $\Sigma_{\mathrm{CSD,max}}$ is the maximum density in the CSD. In
Fig.~\ref{fig:area_vs_t}, we show $A_{\mathrm{gap}}$ as a function of time for
the binaries q1.0e0.0-q1.0e0.8. After an initial transient phase all binaries
settle into a configuration where the $A_{\mathrm{gap}}$ does not change on
longer timescales, but oscillates around a constant value on a timescale $t
\simeq 300\, T_{\mathrm{bin}}$. There is a clear trend of increasing area of
the cavity with increasing eccentricity. While the circular-orbit and
low-eccentricity binaries q1.0e0.0 and q1.0e0.2, and the medium-eccentricity
binary q1.0e0.4 show very similar dynamics, the high-eccentricity binaries
q1.0e0.6 and q1.0e0.8 show a significantly increased area of the cavity. We
note that we do not show radial profiles of surface density $\Sigma / \Sigma_0$
and the area of the cavity $A_{\mathrm{gap}}$ as a function of mass ratio $q$ as
the overall values of these quantities only depend very weakly on $q$ in the
parameter range we have explored ($0.4 < q < 1.0$). One exception is the
timescale of the variability of $A_{\mathrm{gap}}$.

The increased amount of gas inside the cavity and the complicated flow dynamics
for eccentric orbits raises the exciting possibility of the presence of
detectable emission from shocked and unshocked material therein. In order to
investigate this further we show the z-component of the potential
vorticity $\left(\nabla \times \vec{v}\right)^z / (\Sigma/\Sigma_0)$ in
Fig.~\ref{fig:2d_divv} (Li et al. 2005, Dong, Rafikov, \& Stone 2011). We focus
our attention on the non-eccentric binary q1.0e0.0 and the eccentric binary
q1.0e0.6 and display a snapshot in time where the binary is close to apastron
for the eccentric case.  The snapshots correspond to panels (c) and (g) from
Fig.~\ref{fig:2d_rho_1orbit}.  The top row shows surface density $\Sigma /
\Sigma_0$ for comparison, while the bottom row shows $\left(\nabla \times
\vec{v} \right)^z/ (\Sigma/\Sigma_0)$. Large positive values of $\left(\nabla
\times \vec{v} \right)^z/ (\Sigma/\Sigma_0)$ indicate highly compressed and
potentially shocked material.  Compressed/shocked material appears to be
present in the accretion streams onto the CSDs for both the circular and
eccentric orbit case. This traces out the spiral streams in the circular orbit
case and the figure-eight-like structure in the eccentric case highlighting the
possibility of using this feature as a potentially powerful observational probe
of eccentricity. 

\section{Conclusions}

Two-dimensional hydrodynamical simulations have been carried out to investigate
the morphology of gas flows within the cavity region between a central binary and
its CBD in eccentric systems. We find that the streams flowing from the
CBD into the cavity are broad and can take the form of tightly wound streams or a
figure-eight-like structure surrounding the binary for sufficiently eccentric
systems ($e \geq 0.4$), with the specific pattern dependent on the orbital
phase of the system.  This is in contrast to the gas flows characteristic of a
binary in a circular orbit or one of low eccentricity, in which relatively thin
loosely bent streams directly flow from the CBD to the CSDs with the flow
pattern varying little with orbital phase. We have shown that the gas in the
cavity is likely to be shocked and, thus, observational strategies using
molecular lines that are shock tracers are more likely to uncover these
features. 

It is worth asking whether the shock patterns observed in this study would be
present in a fully three-dimensional calculation. Ultimately, 3D calculations
will be necessary to answer this question completely, but it is reassuring that
the characteristic size and shape of the shock structures consist of length
scales significantly larger than the scale height, and therefore the
expectation is that similar structures would be found in 3D, and in particular
the figure-eight--like flow patterns would still be expected to emerge for sufficiently
eccentric orbits.

Thus, the existence of the unique flow patterns in the cavity region
could serve as a proxy to signify a modest to large eccentricity in the orbital
motion of the binary stellar components.  This imprint would serve as
additional support for the inference for an eccentricity in the binary system
in addition to the observations pointing to large disk gap sizes greater than
about 1.7 times the orbital separation of the binary (Artymowicz \& Lubow
1994).  

Although there are orbital phases close to periastron in the case of an
eccentric binary where the morphology of the gas flow in the cavity corresponds
to two streams flowing directly from the CBD to the CSDs,  similar to that in
the case for circular orbit, the diversity of patterns represented throughout
the bulk of the binary orbit indicate that the description of the morphology
for binaries characterized by circular orbits does not necessarily transfer to
binaries in eccentric orbits. Because of this result, the amount of
matter in the cavity varies throughout the orbit with the tendency for a
greater amount to exist in the cavity for binaries characterized by higher
eccentricities. Due to the presence of this matter, the characterization
of the CBD in terms of an inner radius becomes less well defined. We suggest
that the area of the non-emitting region can be an alternative measure.
Given a model for converting the gas surface density to the molecular 
emission, the flux observed from the CSDs can be used to define a
threshold value for the detection/non-detection of emission/gas within the 
cavity as described in Fig. 4. As long as this threshold is chosen to be
greater than the sensitivity for the instrumentational/observational setup it can
be used to determine the area of the cavity as the region with emission below
the threshold.

We have also identified the pattern of the gas flows as a specific
diagnostic to probe the effect of orbital eccentricity, however, it is
challenging to infer the eccentricity of the system since the orbital phase of
the binary system for a given observation is unknown.  In order to provide
further information to place constraints on the eccentricity of the system from
the structure of the gas flows alone, it is necessary to study the kinematics
of the matter in the two streams. Toward this end, detailed comparisons
between observations and theory would be more meaningful if theoretical data
cubes were used to produce synthetic observations taking into account the
detector and telescope simulator for interferometric observations. Future
observational studies are encouraged to carry out high-sensitivity spectral
observations at high spatial resolution to provide a further understanding of
the gas flows in these binary systems in their earliest stages of evolution.

\section*{Acknowledgements}

P.M. and R.E.T. would like to thank Ya-Wen Tang and Ann Dutrey for stimulating
discussions all throughout the process of this work. P.M. thanks Daniel D'Orazio for
discussions regarding eccentric binaries in \texttt{Disco} and analysis of
output data. P.M. acknowledges support by NASA through Einstein Fellowship grant
PF5-160140. R.E.T. acknowledges support from the Theoretical Institute for
Advanced Research in Astrophysics in the Academia Sinica Institute of Astronomy
\& Astrophysics. P.D. acknowledges support from the Institute for Theory and
Computation at the Harvard-Smithsonian Center for Astrophysics. Numerical
calculations were carried out at the UC Berkeley BRC cluster \textit{savio}.


\end{document}